\documentstyle[tighten,preprint,aps,epsfig]{revtex}

\begin{document}

\title{Quark Model study of the NN*(1440) components on the deuteron}

\author{B. Juli\'a-D\'\i az $^{(1)}$, D.R. Entem $^{(1)(2)}$, 
F. Fern\'andez $^{(1)}$, and
A. Valcarce $^{(1)}$}
\address{$(1)$ Grupo de F\' \i sica Nuclear,
Universidad de Salamanca, E-37008 Salamanca, Spain}
\address{$(2)$ Department of Physics, University of Idaho, Moscow, Idaho 83844} 

\maketitle

\begin{abstract}
We present a calculation of the deuteron wave function
including $NN$, $\Delta\Delta$, and $NN^*$(1440) channels.
All the transition potentials, as well as the direct
$NN$ potential, have been obtained from the same
underlying quark model using standard techniques available
in the literature.
The calculated weights for the different
channels are in agreement with
previous theoretical estimations.            
\end{abstract}

\vspace{2cm}

An important effort has been devoted during the last decade
to the quantitative understanding of the $NN$ interaction
in constituent quark models \cite{shimizu,oka,faessler,jpg,val95,david}.
These models consider hadrons as clusters of confined
non-relativistic constituent quarks interacting through 
forces related to QCD. 
The original one-gluon exchange (OGE),
which provided the first explanation for the repulsive 
core of the
$NN$ interaction, was supplemented by Goldstone-boson 
exchanges at the level of quarks when it was realized that 
the constituent quark mass is a signal of the 
spontaneous chiral symmetry breaking ($\chi SB$) producing
fully quark-model based nucleon-nucleon potentials.

Constituent quark models are ideally suited to study the 
influence of baryon resonances on $NN$ phenomenology,
because both, nucleon and baryon resonances, are
described within the same scheme, the parameters 
involved in the calculations being fixed from the
$NN$ sector where a huge amount of data are available.

In the early days of the $NN$ quark-model
calculations it was reasonably established \cite{oka,faessler} 
that the coupling
of $NN$, $\Delta\Delta$ and hidden color
components through the central part of the quark-quark
interaction provided almost negligible
contributions. However, it was soon demonstrated \cite{val95}
that the 
tensor coupling to $N\Delta$ configurations results
of tremendous importance in order to reproduce the low-partial
waves experimental data. It arises therefore the question
of the role of nuclear resonances in the two 
nucleon phenomenology.

This problem has already been undertaken by 
Maeda et al. \cite{maeda}. They have studied
S-wave scattering 
using the resonanting group method (RGM) including not only $N$'s and $\Delta$'s,
but also their excited states,
being the effects of other resonances smaller.
However, due to the naive model they use, they 
are not able to reproduce the deuteron binding energy
and thus, their conclusions should be taken with care.

\section{Baryon-Baryon interactions}
In order to asses the effects of the 
resonances we require 
to start from a quark model that gives
accurate results for the phase-shift and bound state
properties of the $NN$ system.
The model of Ref. \cite{jpg,david}   
fulfills the above requirements and simultaneously gives 
a correct 
description of the low-lying baryon spectrum \cite{gar}.
In this model the primary ingredients of the quark-quark
interaction are the confining potential and the
one-gluon exchange term for the long- and short-range
part of the interaction. 

In the intermediate region between $\chi SB$ scale ($\sim$1 GeV)
and the confinement scale ($\sim$ 0.2 GeV) QCD is formulated 
in terms of an effective theory of constituent quarks 
interacting through the Goldstone modes associated with $\chi SB$
with an effective lagrangian of the form:
\begin{equation}
L_{ch}=g_{ch} F(q^2)\bar{\Psi}(\sigma + i\gamma_5 \vec{\tau}\cdot\vec{\pi})\Psi \, ,
\label{lag}
\end{equation}
where $F(q^2)$ is a monopole form factor
\begin{equation}
F(q^2) =\left [  \Lambda_{\chi}^2 \over \Lambda_{\chi}^2 +q^2 \right ]^{1\over2} \, .
\end{equation}

As the confining force does
not contribute to the baryon-baryon interaction
we do not consider it henceforth.

The form of the interaction derived from Eq. (\ref{lag})  after a
non-relativistic reduction is \cite{david}:
\begin{equation} 
V_{ij}^{PS}(\vec{q}) = - {g_{ch}^2 \over 4 m_q^2 }
{\Lambda_{\chi}^2 \over \Lambda_{\chi}^2 +q^2 } 
{(\vec{\sigma}_i\cdot\vec{q})(\vec{\sigma}_j\cdot\vec{q}) 
\over m_{PS}^2+ q^2 } (\vec{\tau}_i\cdot\vec{\tau}_j) ,
\end{equation}
\begin{equation} 
V_{ij}^{S}(\vec{q}) = -g_{ch}^2 
{\Lambda_{\chi}^2 \over \Lambda_{\chi}^2 +q^2 }
{1\over m_{S}^2+ q^2 } \, .
\end{equation}
being the OGE potential \cite{derujula},
\begin{equation}
V_{ij}^{OGE}(\vec{q})  = \alpha_s (\vec{\lambda}_i\cdot
 \vec{\lambda}_j)
  \left \{ {\pi \over q^2} - {\pi\over 4 m_q^2} \left [
1 + {2\over3} (\vec{\sigma}_i\cdot\vec{\sigma}_j) \right ] 
  + {\pi\over 4 m_q^2} {{[\vec{q}\otimes\vec{q}]^{(2)}\cdot
[\vec{\sigma}_i\otimes\vec{\sigma}_j]^{(2)}} \over q^2}  \right \},
\end{equation}
where the $\lambda$'s are the color Gell-Mann matrices and 
$\alpha_s$ is the strong coupling constant.
The quark model parameters are those of Ref. \cite{david}.

From the constituent quark-model outlined above
the 
$NN$ as well as the $\Delta\Delta$ potentials
have been calculated through the RGM using the 
approach 
explained in detail in \cite{david}. 
Due to the more involved structure of the $N^*$ wave function we
use the Born-Oppenheimer (BO) approximation for the derivation
of the $NN^*$ interaction.
The procedure followed for the calculation of the 
diagonal part of the interaction is described in \cite{mio}, where the 
direct $NN^*$ potential is calculated and some discussion 
about the properties of the partial waves is done. For the 
purpose of our calculation the non-diagonal terms of the $NN^*$ 
potential have been obtained in the same BO approximation.
The definition of the non-diagonal terms is:
\begin{equation}
V_{N N (L \, S \, T) \rightarrow N N^* (L^{\prime}\, S^{\prime}\, T)} (R)
= \xi_{L \,S \, T}^{L^{\prime}\, S^{\prime}\, T} (R) \, - \, \xi_{L \,S \,
T}^{L^{\prime}\, S^{\prime}\, T} (\infty) \, ,  \label{Pot}
\end{equation}
 
\noindent where
 
\begin{equation}
\xi_{L \, S \, T}^{L^{\prime}\, S^{\prime}\, T} (R) \, = \, {\frac{{\left
\langle \Psi_{N N}^{L^{\prime}\, S^{\prime}\, T} ({\vec R}) \mid
\sum_{i<j=1}^{6}{ V_{qq}({\vec r}_{ij})} \mid \Psi_{N N^*}^{L \, S \, T} ({
\vec R}) \right \rangle} }{{\sqrt{\left \langle \Psi_{N N^*}^{L^{\prime}\,
S^{\prime}\, T} ({\vec R}) \mid \Psi_{N N^*}^{L^{\prime}\, S^{\prime}\, T} ({
\vec R}) \right \rangle} \sqrt{\left \langle \Psi_{N N}^{L \, S \, T} ({
\vec R}) \mid \Psi_{N N}^{L \, S \, T} ({\vec R}) \right \rangle}}}} \, .
\end{equation}                       

It is interesting to emphasize at this point that  
the BO approximation has been proved to give results 
of comparable quality to those obtained through the RGM formalism.
For the sake of clarity 
it can be seen that the on-shell properties for the low 
partial waves obtained by means of the BO 
method are of the same quality (within 3$\%$ for energies below 300 MeV)
as those obtained, using the same quark model Hamiltonian, 
making use of the RGM, (see Fig. 1 of Ref. \cite{tripra}).

\section{Deuteron calculation and results}
To calculate the deuteron binding energy, $E_B$, and the 
deuteron wave function we use a method based  
on a discretization of the 
integral Schr\"odinger equation in momentum space.
The eigenvalue 
problem can be written, in a simplified notation:
\begin{equation}
\sum_j [ E_{i}(p_i)\delta_{ij}+V_{ij} -E \delta_{ij} ] \Psi_j = 0 
\label{eq1}
\end{equation}
where $i$ and $j$ run for all the points of the discretized
space and also for the different channels included in the
calculation. $E$ stands for the energy of the system referred to
the $NN$ threshold and $E_{i}(p_i)={p^2\over2\mu_i}+\Delta M_i$ where
$\Delta M_i$ is the mass difference between channel $i$ and the
$NN$ system.
For the above system to have solutions 
different from the trivial one, the following condition
has to be fulfilled,
\begin{equation}
|  E_{i}(p_i)\delta_{ij}+V_{ij} -E \delta_{ij}  | = 0 
\label{fredholm}
\end{equation}
the values $E$ that satisfy Eq. (\ref{fredholm}) are the energies 
of the bound states of the system. Once the energies have been 
found the calculation of the 
wave function can be easily done just by
solving the linear problem of Eq. (\ref{eq1}). The whole problem
has been solved on a Gauss-Legendre mesh of 48 points per channel that 
gives already stable results. 

We include in the calculation
those channels which can couple to 
the quantum numbers of the deuteron, 
$^3S_1^{NN}$ - $^3D_1^{NN}$, $^3S_1^{NN*}$ - $^3D_1^{NN*}$,
$^3S_1^{\Delta\Delta}$ - $^3D_1^{\Delta\Delta}$ - 
$^7D_1^{\Delta\Delta}$ - $^7G_1^{\Delta\Delta}$. 
Note that the 
first relevant channel, from an strictly energetic
point of view, is the $NN^*$ channel as the $N\Delta$ 
system cannot couple to the deuteron.
The most important feature of the quark model
interaction that is relevant here is the tensor part of the 
pseudo-scalar
exchange that is responsible for all the 
non-diagonal parts of the baryonic interactions used 
in the calculation. 
In table 
\ref{table2} we show the results obtained in
our calculation. The deuteron binding energy is fixed to -2.2246 MeV
for all calculations by fine tunning the quark model parameters.
The first thing to be outlined is that the probabilities of 
the $NN^*$ channels are indeed small as compared to the 
$^3S_1^{\Delta\Delta}$ and $^7D_1^{\Delta\Delta}$,
in agreement with the 
estimation of Glozman et al. \cite{glozman}, where they 
studied the spectroscopic factor for $NN^*$ and $\Delta\Delta$ 
channels finding an upper estimate of $<$10$^{-2}$ and
$<$10$^{-3}$, respectively. 
The prediction we obtain is, however, a factor of 10 
lower than that of Rost \cite{rost} where he found 
an estimate of 0.16$\%$ for the $NN^*$ channels in a meson-exchange
calculation. At the same time 
it can be seen that the probability for the Roper channels is 
a factor of 10 bigger 
than the contributions from the 
$^3D_1^{\Delta\Delta}$ and $^7G_1^{\Delta\Delta}$. 
The inclusion of the Roper channels results on a decrease
of the probabilities of the $NN$ channels and a slight 
enhancement of the $^3S_1^{\Delta\Delta}$ channel.
Concerning the other deuteron properties, the main changes are due 
to the inclusion of the $\Delta$ channels, being the Roper
contribution almost negligible.

\begin{table}[t]
\caption{Deuteron wave function ($\%$)}
\begin{tabular}{cc|cc|cccc|ccc}
\multicolumn{2}{c}{$NN$}&\multicolumn{2}{c}{$NN^*$}&
\multicolumn{4}{c}{$\Delta\Delta$}&\multicolumn{3}{c}{}\\
$^3S_1$ & $^3D_1$ & $^3S_1$  & $^3D_1$ &  
$^3S_1$ & $^3D_1$ & $^7D_1$ &
$^7G_1$&$r_m$(fm)&$A_S$(fm$^{-1/2}$)&$\eta$\\
\hline
95.38&4.62&-&-&-&-&-&-&1.976&0.8895&0.0251\\
95.20&4.56&-&-&0.11&0.0035&0.12&0.0063&1.985&0.8941&0.0250\\
95.17&4.53&0.027&0.024&0.13&0.0036&0.12&0.0062&1.986&0.8944&0.0250\\
\end{tabular}
\label{table2}
\end{table}


\begin{thebibliography}{99}
\bibitem{shimizu} K. Shimizu, Rep. Prog. Phys. 52 (1989) 1.

\bibitem{oka} M. Oka, K. Yazaki, Phys. Lett. 90B (1980) 41.

\bibitem{faessler} A. Faessler, F. Fern\'andez, G. L\"ubeck,
and K. Shimizu, Phys. Lett. 112B (1982) 201.

\bibitem{jpg} F. Fern\'{a}ndez, A. Valcarce, U. Straub, A. Faessler,
J. Phys. G 19 (1993) 2013.

\bibitem{val95} A. Valcarce, A. Faessler, F. Fern\'andez,
Phys. Lett. B 256 (1995) 367.

\bibitem{david}  D.R. Entem, F. Fern\'{a}ndez, A. Valcarce,
Phys. Rev. C 62 (2000) 034002.

\bibitem{maeda} I. Maeda, M. Arima, and K. Masutami
Phys. Lett. B 474 (2000) 255.

\bibitem{gar} H. Garcilazo, A. Valcarce, and F. Fernández,
Physical Review C 63 (2001) 035207,  H. Garcilazo, A. Valcarce, and F.
Fernández, submitted to Phys. Rev. C

\bibitem{derujula} A. de R\'ujula, H. Georgi, S.L. Glashow, 
Phys. Rev. D 12 (1975) 147.

\bibitem{mio} B. Juli\'a-D\'\i az, A. Valcarce, P. Gonz\'alez, and F.
Fern\'andez, 
Phys. Rev. C 63 (2001) 024006.

\bibitem{tripra} B. Juli\'a-D\'\i az, F. Fern\'andez, 
A. Valcarce, and J. Haidenbauer, these proceedings.

\bibitem{glozman} L. Ya. Glozman, and E. I. Kuchina,
Phys. Rev. C 49 (1994) 1149.

\bibitem{rost} E. Rost,
Nucl. Phys. A 249 (1975) 510.
 
\end{thebibliography}
\end{document}